\documentclass{egpubl}
\usepackage{eg2022}
 
\ShortPresentation      
\usepackage[T1]{fontenc}
\usepackage{dfadobe}  

\usepackage{cite}  
\BibtexOrBiblatex
\electronicVersion
\PrintedOrElectronic
\ifpdf \usepackage[pdftex]{graphicx} \pdfcompresslevel=9
\else \usepackage[dvips]{graphicx} \fi

\usepackage{egweblnk} 
\usepackage{amsmath}

\title[AvatarGo: Plug and Play self-avatars for VR]%
      {AvatarGo: Plug and Play self-avatars for VR}

\author[Jose Luis Ponton \& Eva Monclús \& Nuria Pelechano]
{\parbox{\textwidth}{\centering Jose Luis Ponton$^1$\orcid{0000-0001-6576-4528},
        Eva Monclús$^1$\orcid{0000-0002-9645-0510} and
        Nuria Pelechano$^1$\orcid{0000-0002-1437-245X}
        }
        \\
{\parbox{\textwidth}{\centering $^1$Universitat Politècnica de Catalunya, Spain}}
}

\begin{document}
 \teaser{
  \vspace*{-0.5cm}
  \includegraphics[width=1.0\linewidth]{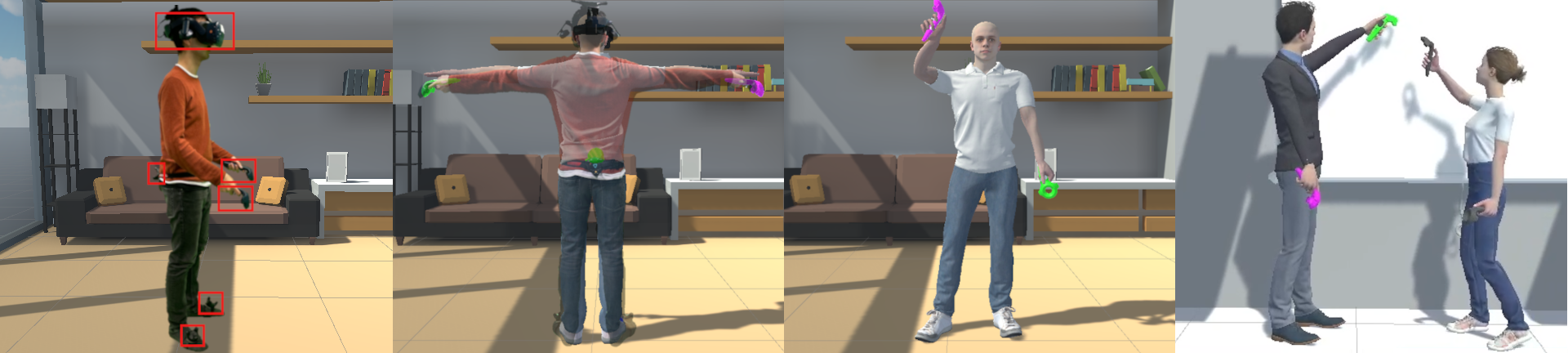}
  \centering
   \caption{Set up for computing the exact offsets between the trackers and the avatar's joints to obtain a better match of poses and movement, using only one HMD, two controllers and three trackers.}
 \label{fig:teaser}
}

\maketitle
\begin{abstract}
The use of self-avatars in a VR application can enhance presence and embodiment which leads to a better user experience. In collaborative VR it also facilitates non-verbal communication. Currently it is possible to track a few body parts with cheap trackers and then apply IK methods to animate a character. However, the correspondence between trackers and avatar joints is typically fixed ad-hoc, which is enough to animate the avatar, but causes noticeable mismatches between the user's body pose and the avatar. In this paper we present a fast and easy to set up system to compute exact offset values, unique for each user, which leads to improvements in avatar movement. Our user study shows that the Sense of Embodiment increased significantly when using exact offsets as opposed to fixed ones. We also allowed the users to see a semitransparent avatar overlaid with their real body to objectively evaluate the quality of the avatar movement with our technique.


\end{abstract} 
\section{Introduction}
The recent pandemic has forced most of the world to work remotely using video conferencing. While this technology allows us to see each other, and facilitates non-verbal communication, the 2D nature interaction is a limiting factor.
A widely shared vision for real collaboration is based on immersive virtual reality (IVR). Several users wearing Head Mounted Displays (HMD) should be able to discuss complex 3D data in a natural manner while sharing the virtual space. This level of collaboration requires the users to be represented with self-avatars that accurately follow their movements. Areas such as architecture, medicine or teaching, could highly benefit from having such enhanced communication through gestures, pointing and interacting with 3D models and with each other. In order to bring this technology main stream, it is essential to facilitate ready to use avatars for any VR application. 
Virtual humanoids can be created using expensive 3D scanners 
or tools that start from a standard avatar model and allow the user to modify its features through a GUI (e.g:  Autodesk Character Generator, MakeHuman, or MetaHuman).
Despite all the recent work to ease the process of generating virtual humanoids, the main difficulty is still to bring these avatars to any IVR application and have the user rapidly set them up and be ready for animation using a few low cost trackers. 

We present \textit{AvatarGo}, self-avatars ready to be incorporated in any VR application with a low cost set-up consisting of a HMD, 2 controllers and 3 trackers. Our system takes the user through a simple and quick set up that scales an avatar and computes the exact offsets between the trackers and the avatar's joints to provide high fidelity movements. 
It simplifies the set up and initiation of trackers, the pairing between trackers and end effectors of the avatar, and it naturally adapts the avatars' hands to the controllers. The user can easily walk into his/her own avatar and with just a few steps, have the avatar accurately following his/her movements.

\section{Related Work} \label{related_work}

When using a HMD, users cannot see their real bodies. A virtual body can replace the participant’s body and if synchronized with the user movement, can induce full-body illusion. This, in turn, enhances the exploration and interaction capabilities of VR \cite{gonzalez-franco_model_2017}. 
The \textit{Sense of Embodiment} (SoE) presented in \cite{kilteni_sense_2012} is the feeling of being inside, controlling, and having a virtual body. It has three sub-components: \textit{Sense of Self-Location} (sensation of being inside the virtual body), \textit{Sense of Agency} (sensation of having control over the virtual body), and \textit{Sense of Ownership} (sensation of the virtual body being one’s body). 
When using avatars in collaborative VR, even a simple representation such as spheres with eyes to indicate the head orientation \cite{andujar_vr-assisted_2018}, or an upper body cartoonish avatar with floating hands can improve non-verbal communication. However, full-body avatars can help users to perform cooperative tasks more accurately and quickly \cite{pan_impact_2017}, and improving animations perceptually by, for example, handling self-contacts can have an impact on embodiment \cite{bovet2018critical}. 
Animating a self-avatar could be done from an ego centric view using fish-eye cameras \cite{xu2019mo} as long as the body is visible from the camera. HMDs typically work with trackers that can be used to 
animate a self-avatar using IK solutions \cite{aristidou2018inverse}, such as IKinema or FinalIK. However, from our experience, previous solutions are not simple to include in any VR project, and suffer from many problems with IDs and orientation matrices. Also, assumptions are made regarding the offsets between trackers and joints that lead to unnatural poses due to the lack of accuracy between the user's body and the avatar. To improve the results, the user has to manually input information or move around trackers through a time consuming and error prone tweaking process.

\section{Self-Avatar construction}

\subsection{Hardware setup}
The system requires one HMD compatible with SteamVR, such as HTC VIVE, two hand controllers and three HTC VIVE Trackers (placed as shown in Figure \ref{fig:teaser} left). Two trackers are placed on the feet and define the ankles’ position, as well as foot orientation. 
The third tracker defines the position of the root, and it is placed on the lower back to minimize the distance between the tracker and the avatar's root joint. We recommend using straps as tight as possible for all three trackers to avoid jittering. The next step consists of finding an exact correspondence between each tracker and the body joint that will be animated following the trackers input.


\subsection{Walk-In-Avatar step} \label{section_avatar_construction}
Users can select any humanoid-like avatar imported in FBX format to Unity. 
After putting on the HMD, users have to stand in T-Pose facing a specific direction and press the trigger button of any controller to start the calibration process. The calibration, consists of fitting a plane through the location of all tracking devices, to then automatically identify each tracker’s role based on their relative location on the plane. 

The HMD height is stored as the user’s eye height and used to scale the avatar uniformly. Then the avatar is rendered in T-pose in the center of the scene facing a virtual mirror and with the feet correctly located on the floor. The user has to walk into the avatar, and find a good alignment between his/her body and the avatar with the help of the virtual mirror. While performing this step, the user can see the avatar and all the tracking devices located on the user's body, therefore the task consists of simply positioning the feet trackers over the avatar's feet, the back tracker on the avatar's lower back, and centering the HMD with the avatar's head. Once this is done, the user presses a trigger to activate the next step.

\subsection{Computing exact offsets}
\begin{figure}[htb]
  \centering
  \includegraphics[width=.8\linewidth]{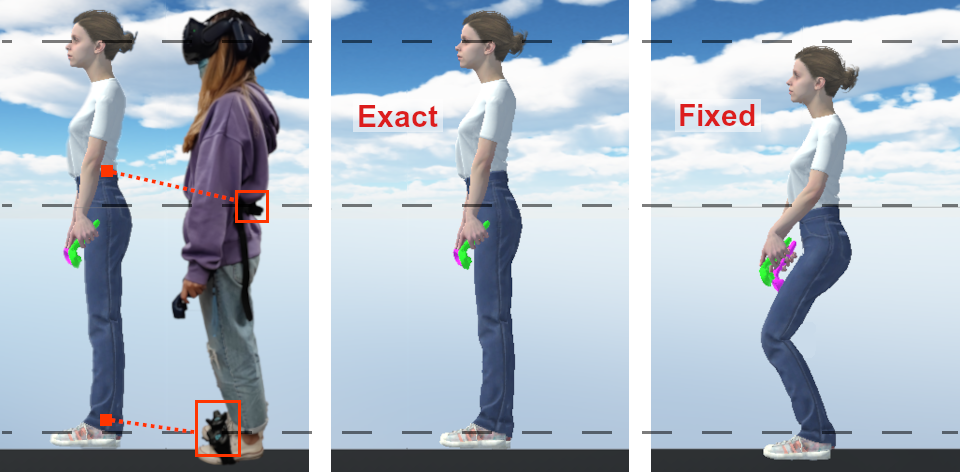}
  \caption{\label{fig:legs_errors}
           A user standing up (left) next to an avatar of user height but longer legs. If exact offsets are computed the avatar pose matches the user (middle). If a fixed offset is assigned between the root joint and the back tracker, the legs appear bent (right).}
\end{figure}

This is a key step, because any mismatch will lead to incorrect positioning of limbs during animation regardless of the IK method used. For example, simply using the back tracker as a root position will give problems with the avatar root appearing a few centimeters behind the user, and with the legs bent if the avatar's legs are longer than the user as shown in Figure \ref{fig:legs_errors}, or floating otherwise.

Each joint has specific limitations and thus needs its own computation to achieve the most accurate animation later on. Therefore, we need to store the initial displacement and rotation between each tracker and the corresponding joint (end effector) that will be used later by the IK to animate the avatar. 


\textbf{Root and feet.} 
We extract 3 displacements between each tracker $T_i$, and its corresponding joint $J_i$, where $i \in \{root, \; right\_foot,\\ left\_foot\}$. For clarity, we remove the subindex $i$ in all equations, since for each body part, $i$ would be replaced in all $J$ and $T$ of the equations, by the same name corresponding to that body part.
While the user is standing in T-pose inside the avatar, we extract the following variables (initiation step, $t=0$):

\begin{itemize}
\item $\mathbf{p}_t(T)$: position of the tracking device $T$.
\item $\mathbf{p}_t(J)$: position of the joint $J$.
\item $\mathbf{v}_t(TJ)$: vector from the tracker position $\mathbf{p}_t(T)$ to the avatar’s corresponding joint position $\mathbf{p}_t(J)$.
\item $\mathbf{R}_t(T)$: tracking device’s $T$ rotation.
\item $\mathbf{R}_t(J)$: joint $J$ rotation.
\end{itemize}


\textbf{Head and back.} 
To position correctly the avatar's head and bend the avatar's back following the head position, we need to store the vector that joins the initial root position with the center of the head: 
\begin{equation}
\mathbf{w}_0 = \mathbf{p}_0(T_{head})-\mathbf{p}_0(T_{root})
\end{equation}

\subsection{Avatar movement}
Unity’s built-in IK was used for the full-body motion due to its versatility and easy-to-use setup. End effectors are defined for the IK, at the wrists and ankles. However it is necessary to also include control for the root and head position, as this is not taken into account by default. Therefore we now describe how each joint is computed in real time based on the trackers' movement.

\textbf{Root and feet.} 
The position of the end effector $\mathbf{p}_t(J)$ is computed in real time using the initial vector offset between the tracker and the joint, $\mathbf{v}_0(TJ)$, the rotation computed at the initiation step, $\mathbf{R}_0(T)$, and the current rotation of the tracker, $ \mathbf{R}_t(T)$:
\begin{equation}
\mathbf{p}_t(J) = \mathbf{p}_t(T) + \mathbf{R}_t(T) \mathbf{R}_0(T)^{-1} \mathbf{v}_0(TJ)
\end{equation}
The current rotation for the end effector (i.e., the target rotation for the joint) is computed as:
\begin{equation}
  \mathbf{R}_t(J) = \mathbf{R}_t(T) \mathbf{R}_0(T)^{-1} \mathbf{R}_0(J) 
\end{equation}
This ensures that the root and ankle joints follow the trackers movement and rotations correctly, regardless of the avatar leg size (thanks to the exact offsets, $\mathbf{v}_0(TJ)$, which are uniquely computed for each user's feet and root). Thus we guarantee that initially the avatar will be standing in a T-pose just like the user, and then all movements will be synchronized (e.g. avatar will bend the knees when the user does, and straighten them exactly as the user does). Our method can handle any initial orientation of the trackers, since we compute the current rotation of the joint, taking into account the initial rotation, $\mathbf{R}_0(T)$. This frees the user from having to follow specific instructions regarding the initial orientation of trackers. 

\textbf{Head and back.} Since Unity IK does not have an end effector for the head, we use forward kinematics to incline the avatar’s spine. 
During simulation time we can compute the angle that rotates the initial root to head vector, $\mathbf{w}_0$, to the current one, $\mathbf{w}_t$. 
\begin{equation}
\alpha_t = \angle (\mathbf{w}_0,\mathbf{w}_t)
\end{equation}
We then apply this angle, $\alpha_t$, to the spine joint of the avatar to bend its back following the user’s movement. Finally, the rotation of the HMD tracker is applied to the head joint to orient the avatar's head.

\textbf{Hands.} For better embodiment, it is essential that the hands appear to be holding the controller at all times to match the user's haptic and visual feedback. 
The wrist joint is computed so that the center of the controller always appears under the palm (keeping an exact offset between the palm and the wrist joint). If the avatar`s arm was shorter than the user's arm, there could be positions for which the avatar's hand could not reach the controller. When those cases are detected, we attach the virtual controller to the hand even if it means not being co-located with the physical controller. 



\begin{figure}[htb]
  \centering
  \includegraphics[width=0.9\linewidth]{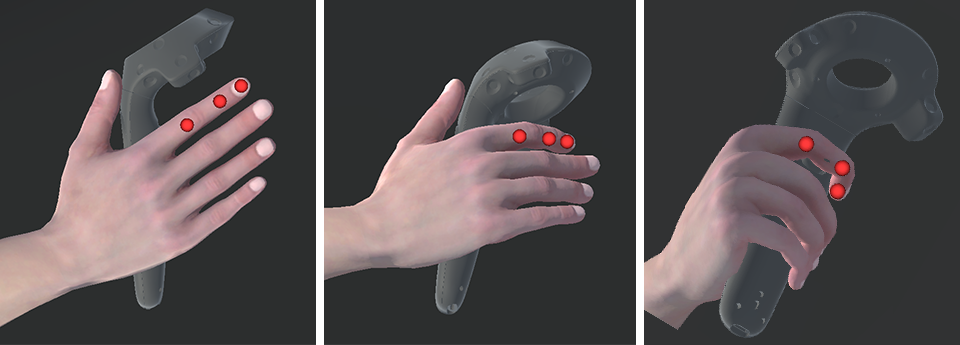}
  %
  \caption{\label{fig:fingersik}
           Fingers are automatically adapted by minimizing their distance to the controller. The three red spheres represent the joints used to perform gradient descent on the index finger.}
\end{figure}

\textbf{Fingers IK.} 
Once the palm is on the controller, the fingers are adapted to hold it correctly following a dedicated IK solution based on gradient descent as shown in Figure \ref{fig:fingersik}.
There are several advantages in using gradient descend for IK. For example, it simplifies adding constraints such as limiting finger rotations. Moreover, several goals can be minimized simultaneously, such as having the thumb minimizing the distance to a specific button of the controller, while minimizing its distance to the surface of the controller. 


The controller can be represented as the signed distance function (SDF) of a capsule, i.e., given a point, the function returns the closest distance to the capsule (negative if the point is inside it). Given two points $\mathbf{s}, \mathbf{e}$ representing the start and end points of the capsule, the radius $r$, and the query point $\mathbf{p}$, we define the capsule SDF as:
\begin{equation}
sdf(\mathbf{p}) = \|\mathbf{v}_1 - \mathbf{v}_2 h\| - r 
\end{equation}
where $h = clamp((\mathbf{v}_1 \cdot \mathbf{v}_2) / (\mathbf{v}_2 \cdot \mathbf{v}_2))$,  
$\mathbf{v}_1 = \mathbf{p} - \mathbf{s} $, and
$\mathbf{v}_2 = \mathbf{e} - \mathbf{s}$.
However, we are not directly moving points, instead, we define the open and close pose of each hand, and for each finger $f \in \{1, \dots, 10\}$ and its joints $j \in \{1, 2, 3\}$ a parameter $t_f^j \in [0, 1]$ that interpolates from the open rotation to the closed one. We define the following vector: $\mathbf{t} = (t_1^1, t_1^2, t_1^3, \dots, t_{10}^1, t_{10}^2, t_{10}^3)$ containing all interpolation factors for all joints. To find the best set of $\mathbf{t}$ we minimize the distance from the capsule to each finger joint. To stop the fingers from getting inside the controller, we multiply the result of $sdf$ by a penalization factor when the distance is negative and then apply the absolute operator.

Let $pos_f^j(\mathbf{t})$ return the position of the joint $j$ of the finger $f$ after applying the interpolation using $\mathbf{t}$. The distance from the capsule to each finger is calculated as:
\begin{equation}
d_f = sdf(pos_f^1(\mathbf{t})) + sdf(pos_f^2(\mathbf{t})) + sdf(pos_f^3(\mathbf{t}))
\end{equation}
We compute the gradient $\nabla d_f$ numerically. Then minimize
$d_f$ updating $t_f^1$, $t_f^2$, $t_f^3$ according to the gradient as:
\begin{equation}
t_f^j = t_f^j - \eta \frac{\partial d_f}{\partial t_f^j}
\end{equation}
where $\eta$ is the learning rate that controls how fast we follow the gradient; a small value will take more time to reach a good solution, while a high value will make the fingers jitter.

\section{User study}
We focus our evaluation on self-avatar animation, because users in VR can easily notice mismatches between the self-avatar and their own movements. Whereas when it comes to collaborators' avatars, animations are less crucial since they cannot see both the collaborators and their avatars simultaneously. To evaluate to what extent our method for calculating exact offsets could enhance embodiment (SoE), we performed a user study to compare the animated self-avatars having fixed offset against our method computing exact offsets. 
We analyzed only the \textit{Sense of Agency} (\textit{SoA}) and the \textit{Sense of Ownership} (\textit{SoO}), since users are always in first-person view and thus \textit{self location} was guaranteed in all cases. 

We performed a within subject experiment with two conditions: (F) Fixed offset and (E) Exact offsets, in random order for each user.  
Each condition was tested first seeing only the Virtual Avatar (VA), and then seeing a semitransparent Overlaid Avatar (OA) with the user body shown using the pass-through camera of the HMD. The goal of the OA was to allow the user to evaluate the self-avatar poses in a more objective manner.
For each condition, the user had to perform three tasks: (T1) \textit{Arms task}, while in T-Pose, move the controllers towards the body and then outwards; (T2) \textit{Legs task}, do squats; (T3) \textit{Free task}, users were encouraged to try any movement.

Before starting the tasks, we gave users some time to get used to the avatar while looking at a mirror. 
We asked six questions: Ag1 for T1, Ag2 for T2, and Ag3 for T3, followed by Ow\{1,2,3\}. Questions were displayed inside VR while doing the exercise to avoid breaking the flow of the experiment.
Users had to score the following questions from 0 (strongly disagree) to 7 (strongly agree): 
\begin{itemize}
\item \textbf{Ag1}. \textit{I felt I was controlling the movement of the virtual arms.}
\item \textbf{Ag2}. \textit{I felt I was controlling the movement of the virtual legs.}
\item \textbf{Ag3}. \textit{The movements of the virtual body felt like they were my movements.}
\item \textbf{Ow1}. \textit{It felt like the virtual body was my body.}
\item \textbf{Ow2}. \textit{It felt like the virtual body parts were my body parts.}
\item \textbf{Ow3}. \textit{The virtual body felt like a human body.}
\end{itemize}


\section{Results}
\begin{figure}[htb]
  \centering
  \includegraphics[width=0.95\linewidth]{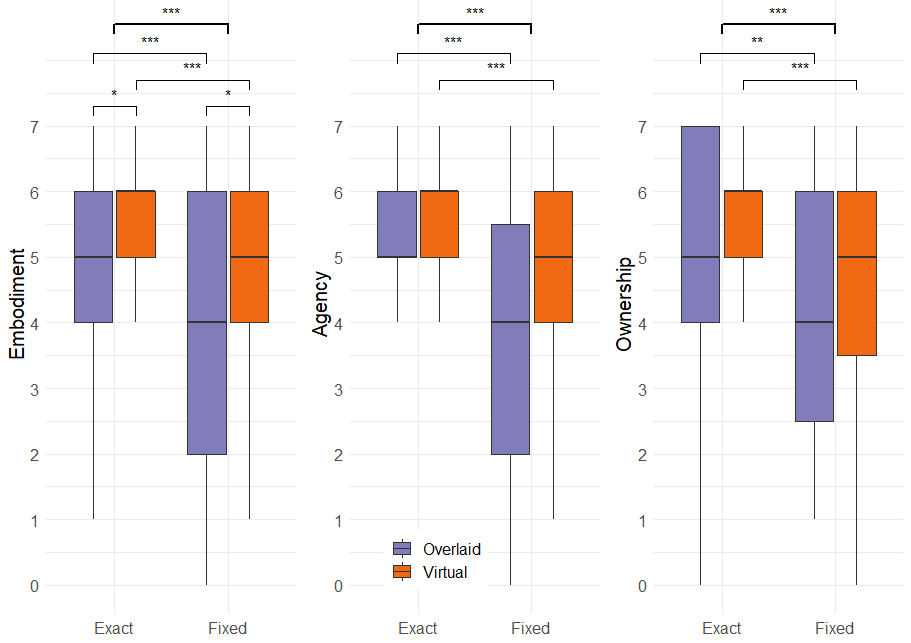}
  %
  
  \caption{\label{fig:boxplot}
           Results of \textit{SoE}, \textit{SoA} and \textit{SoO} comparing exact vs.  fixed offsets, for both render modes: VA and OA. Users rated significantly higher the avatar with exact offsets.}
\end{figure}

A total of 9 participants took part in the experiment (3 females, mean age and standard deviation: 31 $\pm$ 14 years). Most participants reported low to medium experience in VR (2 high, 4 medium, 3 low). 
We tested the normality of the data using a Shapiro test, and since some of the data was not normally distributed, we used the non-parametric Aligned Rank Transform (ART) ANOVA with contrast tests for pairwise comparisons. We also report effect size using Cohen's $d$ statistic for pairwise comparisons. Results on SoE are summarized in Figure \ref{fig:boxplot}.
Users reported significantly higher SoE ($p < .001$, $d=0.60$) for the avatar with exact offsets as opposed to having fixed offsets. 
For both conditions, the SoE was significantly higher when seeing only the virtual avatar (F: $p < .05$, $d=0.56$; E: $p < .05$, $d=0.53$) instead of the overlaid avatar.
Comparing fixed offsets against exact offsets, results of SoE were significantly higher for exact offsets whether they saw only the virtual avatar ($p < .001$, $d=1.00$) or the overlaid avatar ($p < .001$, $d=1.03$).

Overall, users reported high ratings of the SoE for the avatar with exact offsets when using the virtual avatar ($med_{E_{VA}} = 6$ and $IQR_{E_{VA}} = 1$). When comparing agency and ownership questions separately, users also reported high ratings of the SoA ($med_{E_{VA}} = 6$ and $IQR_{E_{VA}} = 1$) and of the SoO ($med_{E_{VA}} = 6$ and $IQR_{E_{VA}} = 1$).  For a more in-depth analysis of SoA and SoO, see Figure \ref{fig:boxplot}.





\section{Conclusions}
In this paper we presented \textit{AvatarGo}, a simple and efficient system to rapidly compute exact offsets between the trackers located on a user, and the joints of a virtual avatar. \textit{AvatarGo} can improve the accuracy of the avatars' movement regardless of the IK method being used. It also provides plausible finger positioning over the controller, which is important to match haptic and visual feedback, thus further enhancing the SoE. The user study shows that exact offsets can enhance SoE. The drop in SoE when seeing the overlaid avatar was higher for the fixed avatar than for our exact avatars, since the mismatch was larger for the avatars with fixed offsets.
The source code is available at \href{https://github.com/UPC-ViRVIG/AvatarGo}{https://github.com/UPC-ViRVIG/AvatarGo}.


\section*{Acknowledgements}
This work was funded by the Spanish Ministry of Economy, Industry and Competitiveness (TIN2017-88515-C2-1-R).

\bibliographystyle{eg-alpha-doi} 
\bibliography{references}


\newpage


\end{document}